\documentclass[aps,prl,twocolumn,showpacs]{revtex4}

\usepackage{epsfig}
\usepackage{graphicx}

\DeclareGraphicsExtensions{.eps,.EPS}

\begin{document}

\title{Spin relaxation and band excitation of a dipolar BEC in 2D optical lattices}
\author{B. Pasquiou, G. Bismut, E. Mar\'echal, P. Pedri, L. Vernac, O. Gorceix and B. Laburthe-Tolra}
\affiliation{Laboratoire de Physique des Lasers, CNRS UMR 7538,
Universit\'e Paris 13, 99 Avenue J.-B. Cl\'ement, 93430
Villetaneuse, France}

\begin{abstract}

 We observe interband transitions mediated by dipole-dipole interactions for an array of 1D quantum gases of chromium atoms, trapped in a 2D optical lattice. Interband transitions occur when dipolar relaxation releases an energy larger than the lattice band gap. For symmetric lattice sites, and a magnetic field parallel to the lattice axis, we compare the measured dipolar relaxation rate with a Fermi Golden Rule calculation. Below a magnetic field threshold, we obtain an almost complete suppression of dipolar relaxation, leading to metastable 1D gases in the  highest Zeeman state.

\end{abstract}


\date{\today}

\maketitle

Collisions and interactions between cold atoms trapped in optical lattices are subject to intense investigations.  These studies are for example relevant for novel strongly correlated quantum phases \cite{who1}, and for metrology \cite{who2}. For most of the experiments to date, focus was made on contact interactions between atoms in the lattice vibrational ground state. Dipole-dipole interactions (DDIs), which have attracted considerable attention, especially since the recent production of Bose Eintein condensates (BECs) with highly magnetic chromium atoms \cite{Cr exp}, introduce important new features in this context, because they can couple atoms in different lattice sites due to their long range character \cite{inguscio}. Here we show that DDIs can be used as well to couple different bands in the lattice, when magnetization changing collisions \cite{note} release an energy overcoming the gap between bands. In addition to their interest for quantum computing, higher lattice orbitals may be used to study novel quantum phases, as suggested in \cite{muller}.

Magnetization changing collisions in a cylindrically symmetric system lead to mechanical rotation in analogy to the seminal work of Einstein and de Haas (EdH) \cite{edh}, as suggested by \cite{edhcoldatom1,edhcoldatom2,edhcoldatom3,edhcoldatom4}. Here, we study dipolar relaxation of spin $S=3$ Cr BECs loaded in a strongly confining 2D optical lattice defining an ensemble of 1D quantum gases in cylindrically symmetric sites.  We observe a threshold as a function of the magnetic field, corresponding to the resonant exchange of a quantum of spin excitation and a quantum of rotational excitation in the lattice. Below the threshold, dipolar relaxation does not release enough kinetic energy to induce the change in angular momentum, and we observe an almost complete suppression of dipolar relaxation for atoms in the highest energy Zeeman state (m$_S$=+3); 1D quantum gases in this state remain well below the degeneracy temperature \cite{1Dregime} for tens of ms. Above threshold, dipolar relaxation occurs, with a transfer of population into excited bands of the lattice. While rotating states in each lattice site are then predicted \cite{edhcoldatom4}, and the observation of such rotation would be a proof of the EdH effect, these rotating states are not reached at our lattice depths due to fast tunneling between sites in excited band states.

In our experiments, almost pure Chromium BECs in the absolute ground state $m_S=-3$ \cite{BECCr}, with 20 000 to 30 000 atoms, are loaded into a 2D optical lattice. The lattice is realized by two independent and orthogonal retro-reflected 1D optical lattices, detuned from each other by 160 MHz. All beams derive from a 532 nm solid state single line laser. As the standing wave is turned on sufficiently slowly (within 15 ms), the BEC is loaded adiabatically in the lowest energy band of the 2D lattice, whose maximal depth in each direction is about 25 $E_R$, where $E_R = \frac{h^2}{2 m \lambda^2}$ is the recoil energy, with $\lambda=532$ nm, and $m$ the mass of $^{52}$Cr atoms. The energy difference in a lattice site between the fundamental and the first excited state, $\omega_L / 2 \pi $=120 kHz, is much larger than the chemical potential, and the motion is frozen along two dimensions. The length of the harmonic oscillator ground state wave-function in the lattice sites $a_L=\sqrt{\hbar/m \omega_L}$ is about 40 nm, and the peak density is raised up to $3\times10^{21}$ m$^{-3}$. As the tunneling time from one lattice site to the next is long (40 ms), we consider that the BEC is split into an array of a few hundred independent 1D quantum gases.

We then reduce the magnetic field $B$, so that the Larmor frequency $\omega_0=g_J \mu_B B / \hbar$ ($g_J$ is the Land\'e factor, $\mu_B$ is the Bohr magneton) is adjusted close to $\omega_L$. The direction of $B$, $\vec{u}_z$, is set parallel to the direction of the 1D tubes, $\vec{u}_L$ ($i.e.$ perpendicular to the two standing waves axes). With a radio-frequency sweep, we transfer the atoms in the $m_S=3$ state: dipolar relaxation then becomes energetically allowed.

\begin{figure}
\centering
\includegraphics[width= 2.8 in]{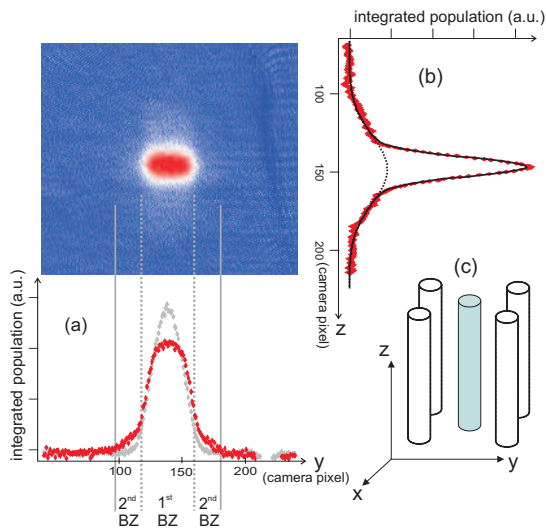}
\caption{\setlength{\baselineskip}{6pt} {\protect\scriptsize
(Color online) Band mapping results above threshold. The false colored picture is an average of 4 absorption images of the BEC released from the lattice. (a) Integrated population profile along $z$: the first excited band (2$^{nd}$ Brillouin Zone, BZ) in the $y$ direction is populated (in red), contrary to below threshold (in grey); noisy data points on the right (due to fringes in the absorption image) were suppressed. (b) Integrated profile along $y$ showing a non Gaussian velocity distribution; lines are results of a double Gaussian fit, yielding an effective temperature. (c) Sketch of the 2D lattice arrangement: the 1D traps are along $z$, the imaging beam is sent along $x$. }}
\label{bandmapping}
\end{figure}

To characterize dipolar relaxation, we implement the following measurement procedure. We measure the number of atoms in the different bands of the lattice by reducing the lattice depth in 200 $\mu$s, which is slow compared to the timescale for band excitation in the lattice ($10$ $\mu$s), but fast compared to the timescale for thermalization (on the order of 1 ms): this band mapping procedure \cite{bandmapping} adiabatically transfers the quasi-momentum distribution in the lattice into the real momentum distribution, which we image after a time of flight equal to 5 ms. The laser beam for absorption imaging being perpendicular to the tubes, we image the population of the different bands of one of the 1D lattice (along $y$), and the velocity distribution along the tubes (axis $z$), as shown in Fig. \ref{bandmapping}.

\begin{figure}
\centering
\includegraphics[width= 2.8 in]{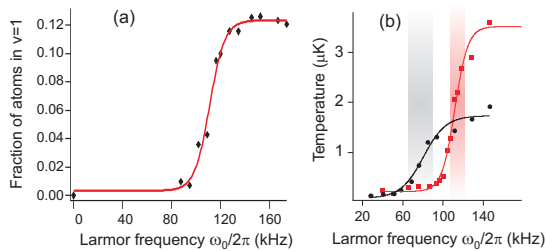}
\caption{\setlength{\baselineskip}{6pt} {\protect\scriptsize
(Color online) Evidence for a threshold in dipolar relaxation: (a) $v=1$ population as a function of Larmor frequency after 25 ms of dipolar relaxation for 25 $E_r$; (b) Effective temperature along the tubes as a function of Larmor frequency after 75 ms, for two different lattice depths (black circles: 12 $E_r$; red squares: 25 $E_r$). The thick grey shaded area and the narrow red shaded area represent the respective widths of the second energy band. Lines are guides for the eye.}}
\label{thresholdv1heat}
\end{figure}

When $B<B_{th}=\hbar \omega_L/(g_J \mu_B)$, we observe almost no heating along the lattice tubes, and no population in higher bands. Typical results for magnetic fields above $B_{th}$ are represented in Fig. \ref{bandmapping}. We observe population in the first excited band ($v=1$), and, within our signal to noise limit, no population in the second one ($v=2$). In addition, we also observe a strong heating of the cloud along the $z$ axis. Population in $v=1$ and the effective (see below) temperature along the tubes are both represented in Fig. \ref{thresholdv1heat} as a function of $B$.

To explain these results, we recall that DDI between two atoms of magnetic moments $\hat{\mu}_1= g_J \mu_B \hat{s}_1$ and $\hat{\mu}_2= g_J \mu_B \hat{s}_2$ ($\hat{s}_{i=(1,2)}$ are spin operators), reads:
\begin{equation}
V_{dd}(\vec{r})=\frac{\mu_0 \left(g_J \mu_B\right)^2}{4 \pi} \frac{\hat{s}_1\cdot\hat{s}_2-3 \left(\hat{s}_1\cdot\vec{u}_r\right)\left(\hat{s}_2\cdot\vec{u}_r\right)}{r^3}
\label{vdd}
\end{equation}
where $\vec{r}=r \vec{u}_r$ is the relative position of the two atoms, and $\mu_0$ is the vacuum permeability. $V_{dd}$ is invariant under simultaneous rotation of spin and space coordinates along any axis, so that the quantum number $m_J=m_S +m_L$ associated to the projection of the total angular momentum $\hat{S}_z+\hat{L}_z$ is conserved in dipolar collisions. In presence of $B$, when $\vec{u}_z \parallel  \vec{u}_L$, and for the specific case of cylindrically symmetric lattice sites, $m_J$ remains a good quantum number: dipolar relaxation results in the creation of rotating states of quantized orbital momentum.

There are two distinct channels of dipolar relaxation when atoms are in $m_S=3$ \cite{dipolarrelaxation}. The first channel corresponds to $\Delta m_S=-1$, the second channel to $\Delta m_S=-2$, and the corresponding release of kinetic energy is respectively $\hbar \omega_0$ and $2 \hbar \omega_0$. Due to conservation of angular momentum, the first (second) channel associated to $\Delta m_L =$ $1$ $(2)$ creates a rotating state of pairs $(x+iy)$ ($(x+iy)^2$), with $x=x_1-x_2$ and $y=y_1-y_2$ the relative coordinates. The corresponding energy costs are respectively $\hbar \omega_L$ and $2 \hbar \omega_L$. As a consequence, both dipolar relaxation channels are expected to become energetically possible for $\omega_0 > \omega_L$, which explains the threshold observed in Fig. \ref{thresholdv1heat}. Dipolar relaxation above threshold therefore involves two  routes, which both populate excited vibrational states in the lattice. Channel 1 populates the first excited lattice band, and, as $(x+iy)^2 = (x_1+ i y_1)^2-2(x_1+ i y_1)(x_2+ i y_2) +(x_2+ i y_2)^2$, channel 2 populates both the first and the second excited bands, and creates (motional) entangled states.

As shown in Fig. \ref{thresholdv1heat}(b), the threshold for dipolar relaxation does depend on the lattice depth, and it corresponds to $\omega_0 > \omega_L =2 \pi \times 120$ $(80)$ kHz for 25 (12) $E_R$. Furthermore, the width of the observed threshold is comparable to the width of the second excited band in the lattice (equal to 16 (43) kHz for 25 (12) $E_R$). As channel 2 is the only one leading to population of the second excited band we deduce that most of our signal comes from it. Finally, Fig. \ref{thresholdv1heat}(b) also shows that raising the lattice depth leads to stronger heating above threshold. Such an increase cannot be solely interpreted as the effect of an increased density (the density only increases by only 40 percent while dipolar relaxation roughly doubles from 12 $E_r$ to 25 $ E_r$), and will be related below to the reduction of the width of the second excited band.

We interpret the strong heating along the tubes above threshold as the consequence of collisional de-excitation (vibrational quenching due to collisions with atoms in $v=0$, see \cite{spielman}). Due to conservation of orbital momentum, vibrational quenching is not possible for rotating states; however,  simple calculations show that a singly or doubly quantized vortex does not survive to a tunneling event. As the tunneling time in the excited bands of the lattice is fast (1 ms in the first band, and 50 $\mu s$ in the second band for 25 $E_r$), atoms rapidly lose their orbital momentum. A symmetry approach shows that $v=2$ atoms can undergo collisional de-excitation with atoms in $v=0$ (contrary to atoms in $v=1$) \cite{spielman}: an atom in $v=2$ can collide with an atom in $v=0$, producing two atoms in $v=0$; therefore population accumulates only in the first excited band, as observed in Fig. \ref{thresholdv1heat}. As the population in $v=1$ rises, vibrational de-excitation between two colliding $v=1$ atoms becomes efficient. Vibrational de-excitation from both bands therefore contributes to the observed heating; this heating is an indirect proof that atoms cease to rotate due to tunneling.

An important feature of our experimental data is that the velocity distribution along the tubes rapidly becomes non-gaussian (see Fig. \ref{bandmapping}). Collisional de-excitation produces pairs of back-to-back moving atoms at a relative momentum set by the lattice depth, so that the created velocity distribution is highly out of thermal equilibrium. It would be expected that it remains so in a pure 1D geometry with only contact interaction, because of the integrability of such systems \cite{weiss}. But in our situation the system is not purely 1D, as the de-excited atoms have enough energy to populate higher bands in the lattice; and DDI may also come into play \cite{DDI integrability}. Despite this lack of integrability, the observed velocity distribution clearly departs from thermal equilibrium even after tens of collisions per particle.  This intriguing lack of thermalization also has an impact on how we  measure heating rates in our system: instead of fitting the velocity distribution along $z$ by a simple gaussian, we evaluate its second moment, hence defining an effective temperature.

\begin{figure}
\centering
\includegraphics[width= 2.8 in]{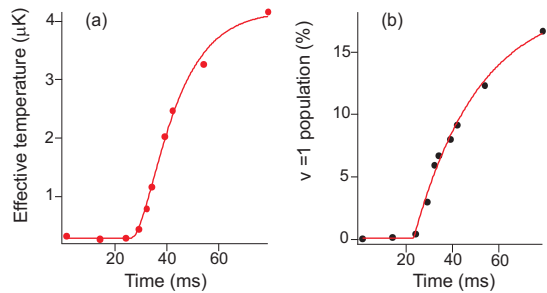}
\caption{\setlength{\baselineskip}{6pt} {\protect\scriptsize
(Color online) Dynamics of dipolar relaxation when the magnetic field is increased above threshold. Threshold is reached at $t$=25 ms, (a) shows the effective temperature (see text), and (b) the measured proportion of atoms in the first excited band. Lines are guides for the eye.}}
\label{dynamicsv1v2}
\end{figure}

With the measured heating rate, and the measured population in $v=1$, we deduce the total rate of kinetic energy deposited in the system by dipolar relaxation above threshold. We plot in Fig. \ref{dynamicsv1v2} the population in $v=1$ and the effective temperature as a function of time. In this experiment, the atoms are first promoted to $m=3$ while $B$ is below the threshold, and then $B$ is raised slightly above the threshold (the threshold is reached at $t=25$ ms in Fig. \ref{dynamicsv1v2}). Below threshold, we observe no heating, and above threshold, we measure an energy increase rate of 215($\pm$30) nK ms$^{-1}$.

To account for these results and estimate a rate parameter, we develop a model describing two particles in a Gaussian trap, assuming that dipolar relaxation can be described by a \textit{local} density dependent rate (as in the 3D and 2D cases, see \cite{dipolarrelaxation}). DDI at threshold couples a two-body state $\Psi_0 \propto \exp\left(-\frac{z^2}{2 z_0^2}\right) \exp\left(-\frac{\rho^2}{2 \rho_0^2}\right)$ describing a pair of atoms in the ground state of a lattice site ($z_0=\left(\frac{2\hbar}{m \omega_z}\right)^{1/2} \gg \rho_0  \equiv \left(\frac{2\hbar}{m \omega_L}\right)^{1/2}$) to rotating states which are excited not only perpendicularly but also along the lattice tubes. In the case of channel 2, the matrix element of DDI writes as $\frac{3Sd^2}{2r^3} \exp\left(2 i \theta \right)$ \cite{dipolarrelaxation}, where $d^2=\frac{\mu_0 \left(g_J \mu_B\right)^2}{4 \pi}$: the ground state is only coupled to even vibrational states along the tubes, corresponding to wave functions $\Psi_2^n \propto \rho^2 \exp(2 i \theta) \exp\left(-\frac{\rho^2}{2 \rho_0^2}\right) \exp\left(-\frac{z^2}{2 z_0^2}\right) H_{2n}\left(\frac{z}{z_0}\right)$ (where $H_n$ stands for the n$^{th}$ Hermite polynomial). Similarly, channel 1 couples to odd vibrational states $\Psi_1^n \propto \rho \exp( i \theta) \exp\left(-\frac{\rho^2}{2 \rho_0^2}\right) \exp\left(-\frac{z^2}{2 z_0^2}\right) H_{2n+1}\left(\frac{z}{z_0}\right)$. As an example, we calculate (for $z_0 \gg \rho_0$):
\begin{eqnarray}
V_1^0 \equiv \left\langle \Psi_0 \left|V_{dd}\right| \Psi_1^0 \right\rangle= 3\left(\frac{\pi}{2} \right)^{1/2}S^{3/2} d^2 \frac{1}{\rho_0^2 z_0} \frac{\rho_0}{z_0} \\
V_2^0 \equiv \left\langle \Psi_0 \left|V_{dd}\right| \Psi_2^0 \right\rangle= \frac{3}{\left(2\pi\right)^{1/2}}S d^2 \frac{1}{\rho_0^2 z_0}
\label{V2V1}
\end{eqnarray}
 We see that $V_1^0\ll V_2^0$, as $z_0 \gg \rho_0$: at threshold, dipolar relaxation is therefore dominated by channel 2. This is consistent with the observed fact that the width of the threshold is comparable to the width of the second excited band of the lattice.

We stress that for our values of the lattice depth, $V_2^0$  (resp. $V_1^0$) is much smaller than the width of the second (first) excited band, $\Delta_2 $ ($\Delta_1$): for 25 $E_R$ (with $\omega_z=2\pi\times400$ Hz), $V_2^0/h\approx 100$ Hz, $V_1^0/h\approx 30$ Hz, while $\Delta_2/h\approx 16$ kHz, and $\Delta_1/h\approx 1.6$ kHz. Thus dipolar relaxation should be described by a Fermi golden rule. It is important though to take into account the coupling of the initial state to all energetically accessible states in the excited bands of the lattice, $i.e.$ to all states $\Psi_i^n$ within an energy band qualitatively set by $\Delta_i$: each of the excited vibrational states (in the i$^{th}$ band) along the lattice axis can be considered as a continuum of states of width $\Delta_i$, and the coupling of the initial state to these many continua can be described as a sum of Fermi golden rules \cite{fano}. In the specific case where channel 2 is dominant this leads to:
\begin{equation}
\Gamma_2= \frac{9\pi}{8} \frac{d^4 S^2 \sqrt{m}}{\rho_0^2 \hbar^2 \sqrt{\Delta_2}} n_0^{3D} N \equiv \frac{1}{2^{3/2}} \beta_2 n_0^{3D} N
\label{beta2}
\end{equation}

Interestingly, the obtained 1D dipolar relaxation rate parameter $\beta_2$ depends on the trapping parameter $\rho_0$, similarly to what we found in 2D \cite{dipolarrelaxation}. In addition, as noticed above, $\beta_2$ depends on the width of the excited band: in that respect, resonant dipolar relaxation at threshold in the lattice is not one-dimensional. For 25 $E_R$, we find $\beta_{2,th} \approx 2 \times 10^{-19}$m$^3$ s$^{-1}$. To relate the measured heating rate of 215 nK.ms$^{-1}$ to a rate parameter, we average over the density of the collection of the 1D gases (see \cite{dipolarrelaxation} for a similar calculation in 2D), and, assuming that dipolar relaxation occurs solely through channel 2, find $\beta_{2,exp} \approx (4.6\pm1.4)\times 10^{-20}$m$^3$ s$^{-1}$. The slight disagreement between $\beta_{2,th}$ and $\beta_{2,exp}$ can be due to the correlations in the 1D quantum gases: in our situation the 1D interaction parameter $\gamma$ \cite{corr} is close to one.

\begin{figure}
\centering
\includegraphics[width= 2.8 in]{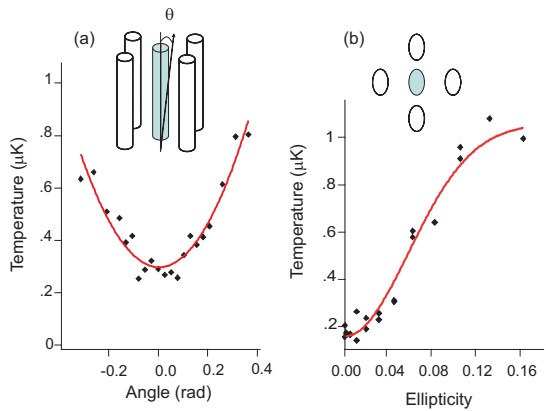}
\caption{\setlength{\baselineskip}{6pt} {\protect\scriptsize
(Color online) Heating in the lattice below threshold after 40 ms, as a function (a) of the angle $\theta$ between the magnetic field and the tube axis, (b) of the transverse ellipticity $(\omega_x-\omega_y)^2/(\omega_x+\omega_y)^2$ of the lattice sites (where $\omega_{x,y}$ are the oscillation frequency of the lattice site along the two main axes $x$ and $y$). Lines are guides for the eye.}}
\label{symmetry}
\end{figure}

Below threshold, we  measure an extremely low heating rate, and we deduce $\beta=(5\pm1.5) \times 10^{-22}$ m$^{3}$ s$^{-1}$. This is typically three orders of magnitude smaller than for 3D degenerate quantum gases \cite{dipolarrelaxation}. As we observe almost no heating for up to 100 ms, this strong reduction enables us to reach a regime where the 1D quantum Bose gas is metastable, despite its highly out-of-equilibrium magnetic state. This important result opens the way to stable spinor mixtures in 1D.

As shown in Fig. \ref{symmetry}, both the cylindrical symmetry of the lattice sites and the proper alignment of the magnetic field along this symmetry axis are needed to reach a regime where dipolar relaxation is strongly suppressed (at low $B$). The observed suppression below threshold is thus directly related to cylindrical symmetry and conservation of angular momentum. Indeed, dipolar relaxation above threshold should populate rotating states in the lattice, in the spirit of the EdH effect. However, due to tunneling in higher lattice bands the rotating states are not produced in our experiment. Using deeper lattices should strongly reduce tunneling and
enable the coherent excitation (when $V_2^0 > \Delta_2$, at typically 75 $E_R$) of such rotating states
in the 2D lattice. Operating in 3D optical lattices in the Mott regime with two atoms per site would also provide a resonant character; as collisional de-excitation in 3D lattices is strongly reduced \cite{muller}, one may reach an intriguing situation where an insulating Mott state in the ground band and a superfluid state in the excited band are coupled by DDIs.

We acknowledge support by Minist\`ere de l'Enseignement Sup\'erieur et de la Recherche (within CPER) and by IFRAF.


\begin{thebibliography}{99}


\bibitem{who1} I. Bloch, J. Dalibard, and W. Zwerger, Rev. Mod. Phys. \textbf{80}, 885 (2008)

\bibitem{who2} C. Lisdat et al., Phys. Rev. Lett. \textbf{103}, 090801 (2009)

\bibitem{Cr exp} J. Stuhler et al., Phys. Rev. Lett. \textbf{95}, 150406 (2005); T. Lahaye et al., Phys. Rev. Lett. \textbf{101}, 080401 (2008); G. Bismut et al., Phys. Rev. Lett. \textbf{105}, 040404 (2010)

\bibitem{inguscio} M. Fattori et al., Phys. Rev. Lett. \textbf{101}, 190405 (2008)

\bibitem{note} Also responsible of demagnetization cooling, see M. Fattori et al., Nature Physics \textbf{2}, 765 (2006).

\bibitem{muller} T. Muller, S. Folling, A. Widera and I. Bloch, Phys. Rev. Lett. \textbf{99}, 200405 (2007)

\bibitem{edh} A. Einstein and W. J. de Haas, Verh. Dtsch. Phys. Ges. \textbf{17}, 152 (1915)

\bibitem{edhcoldatom1} Y. Kawaguchi, H. Saito, and M. Ueda, Phys. Rev. Lett. \textbf{96}, 080405 (2006)


\bibitem{edhcoldatom2} L. Santos and T. Pfau, Phys. Rev. Lett. \textbf{96}, 190404 (2006)


\bibitem{edhcoldatom3} K. Gawryluk, M. Brewczyk, K. Bongs, and M. Gajda, Phys. Rev. Lett. \textbf{99}, 130401 (2007)


\bibitem{edhcoldatom4} B. Sun and L. You, Phys. Rev. Lett. \textbf{99}, 150402 (2007)

\bibitem{1Dregime} D. S. Petrov, G. V. Shlyapnikov, and J. T. M. Walraven, Phys. Rev. Lett. \textbf{85}, 3745 (2000)

\bibitem{BECCr} Q. Beaufils et al., Phys. Rev. A \textbf{77}, 061601 (2008); G. Bismut et al., Appl. Phys. B, Online First (2010)

\bibitem{bandmapping}  A. Kastberg et al., Phys. Rev. Lett. \textbf{74}, 1542 (1995); M. Greiner, I. Bloch, O. Mandel, T. W. Hansch and T. Esslinger, Phys. Rev. Lett. \textbf{87}, 160405 (2001)


\bibitem{dipolarrelaxation} B. Pasquiou et al. Phys. Rev. A \textbf{81}, 042716 (2010)

\bibitem{spielman} I. B. Spielman et al., Phys. Rev. A \textbf{73}, 020702 (2006)

\bibitem{weiss} T. Kinoshita, T. R. Wenger and D. S. Weiss, Nature \textbf{440}, 900 (2006)

\bibitem{DDI integrability} DDIs between tubes can make the system effectively 3D, see D. W. Wang, M. D. Lukin, and E. Demler, Phys. Rev. Lett. \textbf{97}, 180413 (2006).

\bibitem{fano} U. Fano, Phys. Rev. \textbf{124}, 1866 (1961)

\bibitem{corr} $\gamma$ characterizes the crossover between strong ($\gamma \gg 1$), and weak ($\gamma \ll 1$) interactions, see D. M. Gangardt and G. V. Shlyapnikov,  Phys. Rev. Lett. \textbf{90}, 010401 (2003).



\end{thebibliography}
\end{document}